# Effect of Cloud Based Learning Management System on The Learning Management System Implementation Process:

Faculty and Student Perspectives

Ajayi Ekuase-Anwansedo
Science and Math Education
Southern University and A&M, Baton Rouge, Louisiana
United States, ajayi_anwansedo_00@subr.edu

Akai Smith
Office of Admissions and Recruitment
Southern University and A&M, Baton Rouge, Louisiana
United States, akai_smith@subr.edu

**ABSTRACT**

The concept of E-learning in Universities has grown rapidly over the years to include not just only a learning management system (LMS) but also tools initially not designed for learning such as Facebook and advanced learning tools, for example games, simulations and virtualization. As a result, Cloud-based LMS is being touted as the next evolution of the traditional LMS. It is hoped that Cloud-based LMS will resolve some of the challenges associated with the traditional LMS implementation process. In a previous study [10], we reported that lack of involvement of faculty and students in the LMS implementation process results in the limited use of the LMS by faculty and students. The question then is, "Will the cloud-based LMS resolve these issues? We conducted a review of literature and presented an overview of the traditional LMS, cloud computing and the cloudbased LMS and we described how the cloud computing LMS resolve issues raised by faculty and students. we find that even though, cloud-based LMS resolve most of the technical issues associated with the traditional LMS, some of the human issues were not resolved. We hope that this study draws attention to non-technical issues associated with the LMS implementation process.

## 1 INTRODUCTION

The Learning management system (LMS) is a classroom management, communication and collaboration tool used in higher education to facilitate classroom and online learning activities. However, innovations in teaching and learning caused by the rapid development in web technologies [20], the use of social software in learning [8] and advanced learning tools [9] has led to inadequacies in the traditional LMS. Thus, the LMS infrastructure must evolve to incorporate various learning media, informal and formal learning environments and new methods of storing and sharing information to accommodate today's advancement in teaching and learning [12].

The new LMS infrastructure represents what is called the cloud-based LMS, which extends the capabilities of the traditional LMS by integrating it with cloud computing infrastructure. The cloud-based LMS provides more sophisticated tools to protect data from unauthorized access and corruption. it is cost effective in terms of reducing the cost of computing services and the cost of purchasing and maintaining the software and hardware devices [16]. This is possible because users are charged only for the computing services, they use and because the software, and hardware devices resides in the vendor's datacenter [2].

Cloud-based LMS increases compatibility and interoperability amongst computing and technology devices e.g. Third-party devices like cameras, 3D printers, gaming consoles and simulation tools [14]. And promotes ubiquitous access to computing services -from anybody, via any network connection to any computing service, which greatly improves communication and collaboration [2]. Furthermore, the cloud-based LMS allows for customization – enables users adjust computing services based on their preference and is scalable – increases or reduces computing resources based on needs of the user [16, 2, 14].

Even though cloud based LMS has privacy, security. legal, third party dependence and reliability issues [20, 5] which can significantly influence student's decision to use cloud-based LMS [3] It is still a viable eLearning solution because it resolves the frustrations, faculty and students experience with the traditional LMS implementation process. To explore the effect of the cloud-based LMS implementation process, this study is organized as follows. In Section 2, we present the background of the study, exploring the concepts of the traditional LMS, the cloud computing infrastructure and the cloud based LMS concepts. In section 3, we addressed students and faculty concerns about the traditional LMS respectively. In section 4, we presented the conclusion and discussed future research.

## 2 BACKGROUND

### 2.1 Traditional Learning Management System (LMS)

According to [17], an LMS is an online software program which serves as a repository for learning resources such as electronic textbooks, and lecture slides, facilitates learning activities and supports communication and collaboration between faculty and student. Additionally, LMS is used for classroom and course management [17]. Some advantages of using LMS includes tracking of all activities involved in the learning process, rapid creation and distribution of course content and enabling students to learn independently [6].Usually, the LMS infrastructure, which includes a

Ajayi Ekuase-Anwansedo and Akai Smith

licensed software for example, Blackboard and Canvas or an open source solution like Moodle, and the hardware components is built from scratch and maintained by the University.

### 2.2 Cloud Computing

The National Institute of Standards and Technology (NIST) defines cloud computing as follows:

"A model for enabling ubiquitous, convenient, ondemand network access to a shared pool of configurable computing resources (e.g., networks, servers, storage, applications, and services) that can be rapidly provisioned and released with minimal management effort or service provider interaction."[15].

The main characteristics of cloud computing are ondemand self-service, Broad network access, Resource pooling, Rapid elasticity, and Measured service [15].

Basically, cloud computing is an internet-based computing technology which allocate computing resources as a service to various users as they are needed [19, 11]. These resources include software applications (software as a service (SAAS)), the application development environment (platform as a service (PAAS) and computing resources (Infrastructure as a service (IAAS)) [21]. Consequently, the organization or user do not have to locally own or maintain hardware or software resources needed for their work. Usually users utilize these resources (SAAS, PAAS and IAAS) via three models called the deployment models - community, public, private and hybrid cloud [15].

In the community cloud, computing services, are provided and maintained by a service provider such as Amazon through Amazon web services (AWS) exclusively for a community of users with shared values [15]. Public clouds are provided for public users. Private clouds are dedicated to a single organization and the hybrid cloud is a combination of two or more of the deployment models described above.

Figure 1 below depicts an overview of the concept of cloud computing.

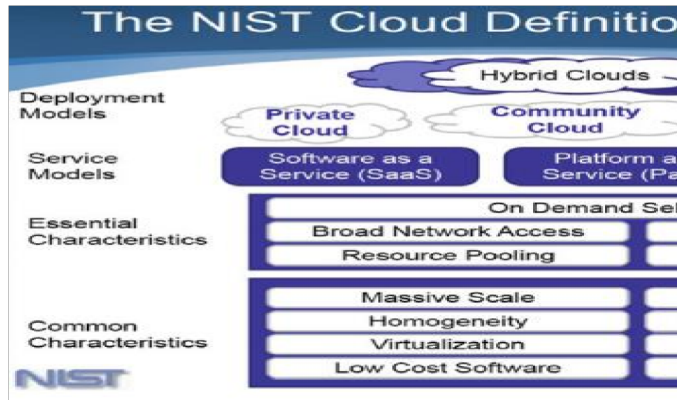

**Figure 1: NIST concept of Cloud Computing adopted from [15, 22]**

### 2.3 Cloud Based LMS

Cloud based LMS is the integration of the LMS and cloud computing. LMS systems takes advantage of the cloud computing infrastructure to provide advanced teaching and learning tools and strategies. Usually, the new LMS provides an environment which consists of an LMS, Social software tools (wikis, blogs) and advanced learning tools like virtual reality and simulations, for users to select tools to support their learning process. An example is the Learning Apps project described by [4]. The Learning Apps project provides an environment where users can personize their e-learning environment using cloud-based infrastructure they select their desired learning tools. Similarly, [20] describes how cloud computing enables the inclusion of social media, wikis, podcasting and weblogs in the e-learning process as a new way for communication and collaboration, between faculty and students and amongst students. [7] developed a Cloud Elearning for Mechatronics (CLEM) for Vocational, Education and Training (VET) teachers. The CLEM tool enables teachers and students connect to laboratories, form a community of teachers and students, and share various mechatronic devices such as 3D printer and stepper motors remotely.

## 3 ISSUES WITH TRADITIONAL LMS IMPLEMENTATION PROCESS

### 3.1 Student Issues

*Inclusion in the LMS Implementation Process* . One of the foremost issues raised in the LMS implementation process is  the non-inclusion of students input in the choice of LMS and the lack of consideration for students needs in the process [10, 12]. In cloud-based LMS, the primary focus of the system is to enable users control their learning environment and learning process. Thus, the student's learning needs are prioritized from the beginning of the implementation process.

*LMS Acceptance and Use* . Another issue resulting from the non-involvement of student in the implementation of the LMS is that of user acceptance. Students use traditional LMS, especially in the classroom, when they are required to. However, this is not the case with Cloud based LMS, because of the user engagement from the beginning of the implementation process, students embrace and willingly use the LMS. Several studies show that students prefer using cloud based LMS services when compared to traditional LMS [13].

### 3.2 Faculty issues

*Time* . According to [1], most faculty will not use a LMS because of perceived time investment in adopting the LMS Time spent transferring the course materials from one LMS application to the other. The Cloud based environment reduces the time taken to transfer course materials from the local computer to the cloud based LMS and vice versa by synchronizing the files in the local computer with the cloud based LMS files [20] [14].

*Lack of understanding of how to implement and use LMS in their classroom - Training* . Faculty often raise the issue of lack of understanding of how to implement LMS in their classroom [1]. Most cloud based LMS services like Facebook, Dropbox and wikis have several training resources developed by the respective companies and a community of users of these tools who provide adequate training and support to users.

*Compensation – Administrative issue* . Faculty mentioned Compensation for their involvement in the LMS implementation process [18], Financial compensation cannot be directly addressed by the cloud-based LMS. The project team or the University responsible for the implementation process decide whether to provide such compensation.

## CONCLUSION AND FUTURE RESEARCH

The cloud-based learning management system represents an improvement on the traditional LMS to include the cloud computing infrastructure. Although, Cloud based LMS addresses some of the issues associated with the LMS implementation process such as inclusion of the student in the implementation process, student acceptance of the LMS and reducing the time to transition from one LMS to the other. Other issues such as compensation for faculty for 

their involvement in the LMS implementation process cannot be addressed directly by the cloud-based LMS process. Therefore, there is a need to address the technological issues as well as the other issues associated with the LMS implementation process.

Most of the paper reviewed on cloud-based LMS presents the student's perspective, acceptance and use of cloud computing services. Future research should be directed towards cloud-based LMS implementation process from the perspectives of the faculty, the technology team and the University leadership to provide a holistic view of the implementation process of the cloud-based LMS in Universities.

## References

[1] Ackerman, D., Chung, C., & Sun, J. C. Y. (2014). Transitions in classroom technology: Instructor implementation of classroom management software. Journal of Education for Business, 89(6), 317-323.
[2] Aldheleai, H. F., Bokhari, M. U., & Alammari, A. (2017). Overview of cloudbased learning management system. International Journal of Computer Applications, 162(11).
[3] Arpaci, I., Kilicer, K., & Bardakci, S. (2015). Effects of security and privacy concerns on educational use of cloud services. Computers in Human Behavior, 45, 93-98.
[4] Alier, M., Mayol, E., Casañ, M. J., Piguillem, J., Merriman, J. W., Conde, M. Á., et al.(2012). Clustering projects for elearning interoperability.Journal of Universal Computer Science, 18(1), 106–122.
[5] Arpaci, I. (2016). Understanding and predicting students' intention to use mobile cloud storage services. Computers in Human Behavior, 58, 150-157.
[6] Caglar, S (2018) The Evolution and Diffusion of Learning Management Systems: The Case of Canvas LMS. In Correia, A.-P. (2018) (Ed.). Driving Educational Change: Innovations in Action (pp. 86-102). eBook, available at https://ohiostate.pressbooks.pub/drivechange/
[7] Chao, K. M., James, A. E., Nanos, A. G., Chen, J. H., Stan, S. D., Muntean, I., ... & Cooper, J. (2015). Cloud E-learning for Mechatronics: CLEM. Future Generation Computer Systems, 48, 46-59.
[8] Dalsgaard, C. (2006). Social software: E-learning beyond learning management systems. European Journal of Open, Distance and e-learning, 9(2).

[9] Dodd, B. J., & Antonenko, P. D. (2012). Use of signaling to integrate desktop virtualreality and online learning management systems.Computers and Education,59(4), 1099–1108.

[10] Ekuase-Anwansedo, A., Craig, S. F., & Noguera, J. (2018,). How to Survive a Learning Management System (LMS) Implementation? A Stakeholder Analysis Approach. In Proceedings of the 2018 ACM on SIGUCCS Annual Conference (pp. 165-168). International World Wide Web Conferences Steering Committee.

[11] Fox, A., Griffith, R., Joseph, A., Katz, R., Konwinski, A., Lee, G., ... & Stoica, I. (2009). Above the clouds: A berkeley view of cloud computing. Dept. Electrical Eng. and Comput. Sciences, University of California, Berkeley, Rep. UCB/EECS, 28(13), 2009.

[12] García-Peñalvo, F. J., Conde, M. Á., Alier, M., & Casany, M. J. (2011). Opening learning management systems

[13] to personal learning environments. Journal of Universal Computer Science, 17(9), 1222–1240.

[14] Hunsinger, D. S., & Corley, J. K. (2012). An examination of the factors influencingstudent usage of dropbox, a file hosting service.In Proceedings of the conferenceon information systems applied research(Vol. 2167, p. 1508).

[15] Karim, I., & Goodwin, R. (2013). Using cloud computing in e-learning systems. learning, 1, 2.

[16] Mell, P., & Grance, T. (2011). The NIST definition of cloud computing.

[17] Muniasamy, V., Ejalani, I. M., & Anandhavalli, M. (2014). Moving towards virtual learning clouds from traditional learning: Higher educational systems in India. International Journal of Emerging Technologies in Learning (iJET), 9(9), 70-76.

[18] Rößling, G., Joy, M., Moreno, A., Radenski, A., Malmi, L., Kerren, A., ... & Oechsle, R. (2008). Enhancing learning management systems to better support computer science education. ACM SIGCSE Bulletin, 40(4), 142-166.

[19] Ryan, T. G., Toye, M., Charron, K., & Park, G. (2012). Learning management system migration: An analysis of stakeholder perspectives. The International Review of Research in Open and Distributed Learning, 13(1), 220-237.

[20] Sarrab, M., Alalwan, N., Alfarraj, O., & Alzahrani, A. (2015). An empirical study on cloud computing requirements for better mobile learning services. International Journal of Mobile Learning and Organisation, 9(1), 1-20.

[21] Scerbakov, A., Ebner, M., & Scerbakov, N. (2015). Using cloud services in a modern learning management system. Journal of computing and information technology, 23(1), 75-86.

[22] Shiau, W. L., & Chau, P. Y. (2016). Understanding behavioral intention to use a cloud computing classroom: A multiple model comparison approach. Information & Management, 53(3), 355-365.

[23] Weber, A. S. (2013). Cloud computing in education. In Ubiquitous and mobile learning in the digital age

[24] (pp. 19-36). Springer, New York, NY.